\documentclass[12pt]{article}
\usepackage{amsmath}
\usepackage{graphicx,psfrag,epsf}
\usepackage{enumerate}
\usepackage{natbib}
\usepackage{url} % not crucial - just used below for the URL 

\usepackage{amsfonts}
\usepackage{multicol, booktabs, bbm, color, xcolor}
\usepackage{subcaption}
\graphicspath{{./img/}}
\newcommand{\blind}{1}

\begin{document}

\def\spacingset#1{\renewcommand{\baselinestretch}%
{#1}\small\normalsize} \spacingset{1}

%%%%%%%%%%%%%%%%%%%%%%%%%%%%%%%%%%%%%%%%%%%%%%%%%%%%%%%%%%%%%%%%%%%%%%%%%%%%%%

\if1\blind
{
  \title{\bf Heterogeneous Distributed Lag Models to Estimate Personalized Effects of Maternal Exposures to Air Pollution}
  \author{Daniel Mork\\
Department of Biostatistics\\ 
Harvard T.H. Chan School of Public Health\\\\
Marianthi-Anna Kioumourtzoglou\\
Department of Environmental Health Sciences\\
Columbia University Mailman School of Public Health\\\\
Marc Weisskopf\\
Department of Environmental Health\\ 
Harvard T.H. Chan School of Public Health\\\\ 
Brent A Coull\\
Department of Biostatistics\\
Harvard T.H. Chan School of Public Health\\\\ 
Ander Wilson\\ 
Department of Statistics\\ Colorado State University}
  \maketitle
} \fi

\if0\blind
{
  \bigskip
  \bigskip
  \bigskip
  \begin{center}
    {\LARGE\bf Heterogeneous Distributed Lag Models to Estimate Personalized Effects of Maternal Exposures to Air Pollution}
\end{center}
  \medskip
} \fi

\clearpage
%\bigskip
\begin{abstract}
Children's health studies support an association between maternal environmental exposures and children's birth outcomes. A common goal is to identify critical windows of susceptibility--periods during gestation with increased association between maternal exposures and a future outcome.  The timing of the critical windows and magnitude of the associations are likely heterogeneous across different levels of individual, family, and neighborhood characteristics. Using an administrative Colorado birth cohort we estimate the individualized relationship between weekly exposures to fine particulate matter (PM$_{2.5}$) during gestation and birth weight. To achieve this goal, we propose a statistical learning method combining distributed lag models and Bayesian additive regression trees to estimate critical windows at the individual level and identify characteristics that induce heterogeneity from a high-dimensional set of potential modifying factors. We find evidence of heterogeneity in the PM$_{2.5}$–birth weight relationship, with some mother-child dyads showing a 3 times larger decrease in birth weight for an IQR increase in exposure (5.9 to 8.5 $\mu g/m^3$ PM$_{2.5}$) compared to the population average. Specifically, we find increased vulnerability for non-Hispanic mothers who are either younger, have higher body mass index or lower educational attainment. Our case study is the first precision health study of critical windows.

\end{abstract}

\noindent%
{\it Keywords:}  environmental health, critical windows of susceptibility, distributed lag models, effect heterogeneity, Bayesian additive regression trees
\vfill

\newpage
\spacingset{1.45} % DON'T change the spacing!

%======================================
\section{Introduction}\label{sec:intro}
%======================================
A growing body of research has found maternal exposure to environmental chemicals during pregnancy to be associated with changes in children’s birth and health outcomes. Detrimental outcomes linked to increased environmental exposures include decreased birth weight \citep{Bell2007AmbientMassachusetts}, increased risk of preterm birth \citep{Stieb2012}, increased risk of asthma \citep{Lee2017,Bose2017PrenatalSex}, and altered neurological outcomes \citep{Chiu2016PrenatalAssociations}, among others \citep{Sram2005AmbientLiterature}. Recently, research has focused on determining the time periods during fetal development, or critical windows of susceptibility, when increased exposure can alter future health outcomes. Identification of critical windows gives insight into how biological processes involved in fetal development may be impacted by exposure to environmental chemicals \citep{Wright2017EnvironmentHealth}. However, exposure effects, including critical window timing and effect magnitude, are likely to vary across a population. Effect heterogeneity may be governed by biological (e.g., sex of child), socioeconomic (e.g., maternal income), or other non-chemical environmental factors (e.g., neighborhood characteristics). 

We aim to identify populations with increased vulnerability to pollution exposure and estimate individualized critical windows and exposure-time-response functions. Estimating individualized critical windows and exposure-response relationships will better inform precision environmental health interventions and bring attention to vulnerable populations, which the Environmental Protection Agency (EPA) is required to protect through the 2016 update to the Toxic Substances Control Act \citep{Krimsky2017TheAct}. Several epidemiological studies have previously considered the question of identifying subgroup specific critical windows for a small number of predefined groups. \citet{Lee2017} and \citet{Bose2017PrenatalSex} found increased risk of asthma due to prenatal exposures to fine particulate matter (PM$_{2.5}$) and nitrate for a subgroup of boys concurrently exposed to high prenatal stress. \citet{Chiu2017PrenatalAssociations} identified early- to mid-pregnancy critical windows where increased PM$_{2.5}$ exposure during pregnancy was associated with increased body mass index (BMI) z-scores and fat mass in boys and waist-to-hip ratios in girls. However, no paper has allowed for a data-driven approach to heterogeneity by simultaneously looking across multiple modifiers. 

%Using an administrative database of 310,236 birth vital statistics records from Colorado, USA we estimate the relationship between PM$_{2.5}$ exposure during pregnancy and resulting birth weight. Specifically, we seek to determine if critical windows of susceptibility and the exposure-response relationship vary across the population and which factors are responsible for this heterogeneity. Our dataset contains a range of individual covariates, including mother's age, weight, height, income, education, marital status, prenatal care habits, smoking before and during pregnancy, as well as race and Hispanic designations. To answer our question, we propose a Bayesian regression tree framework for estimating critical windows and identifying susceptible populations.

A commonly applied method to identify perinatal critical windows and estimate the exposure-time-response relation between maternal exposures and an outcome is the distributed lag model (DLM). In a DLM, an outcome is regressed on repeated measures of an exposure assessed over a time period prior to the outcome. The vector of regression coefficients for linear effects at each exposure time point is referred to as the distributed lag function and characterizes both critical windows and the exposure-response association. Current methods to estimate DLMs that vary across a population are limited to predefined strata or spatially varying effects. In the age of big data, we have access to a host of potential modifiers, but the true modifiers responsible for differences in distributed lag function are unknown. Harnessing this information can lead to personalized environmental health decisions and directly address EPA's mandate to establish regulations protecting populations most vulnerable to exposure.

In this paper we analyze an administrative Colorado birth cohort dataset. We leverage rich data on potential maternal and child modifying factors to answer the following questions. Which individual characteristics modify the PM$_{2.5}$---birth outcome relationship? Do critical windows differ among vulnerable populations? Can we characterize individual-specific exposure-time-response relationships? We explore a range of potential modifiers of the relationship between exposure and birth outcome including: mother's age, body mass index, income, education, smoking habits, marital status, prenatal care, race and Hispanic indicators, and the child's sex. Linking weekly exposure to PM$_{2.5}$ during gestation at a mother's census tract of residence allows insight into the aforementioned questions surrounding critical windows and vulnerable populations. To explore possible effect heterogeneity in the distributed lag function, we propose structured machine learning approaches that combine state-of-the-art regression tree frameworks for both distributed lag function and effect heterogeneity estimation. In a comprehensive simulation we show the potential for bias when applying traditional DLMs in the presence of effect modification and demonstrate how our methods both overcome this bias and allow for in-depth inference through variable selection techniques and individual- or subgroup-specific posterior analyses.

Our data analysis is the first study to estimate individualized critical windows to air pollution exposure and a birth outcome. While a traditional DLM method finds an inter-quartile range (IQR) increase in PM$_{2.5}$ (5.9 to 8.5 $\mu g/m^3$) results in an approximate decrease in birth weight of 11.3g, our heterogeneous analysis estimates the impact to be as much as three times larger among the most vulnerable populations. Among the modifiers considered, our analysis finds that age, BMI, Hispanic designation, and education are potentially important modifiers that were selected as a source of heterogeneity. In particular, the results indicate non-Hispanics with increased BMI are more vulnerable to PM$_{2.5}$ exposures and early- and late-gestation are potential critical windows. In addition, we find individual variability within subgroups due to other modifying characteristics. Software to replicate our simulation and method for applications is available in the \texttt{R} package \texttt{dlmtree}.

\section{Colorado Public Health Birth Records Data}\label{sec:co_birth_data}
%======================================
We acquired an administrative Colorado birth records dataset from Colorado Department of Public Health and Environment. This birth cohort includes all births from Colorado with estimated conception dates between January 1, 2007 and December 31, 2015. Each birth record includes the mother's census tract of residence and a range of individual covariates including: maternal age, weight and height (which we have used to calculate body mass index), income, education, marital status, prenatal care habits, smoking before and during pregnancy, race and Hispanic designations and child sex. Based on the mother's census tract of residence, we assessed PM$_{2.5}$ exposures and temperature during gestation predicted at high spatial and temporal resolution. Specifically, daily PM$_{2.5}$ measurements were obtained from EPA community multiscale air quality modeling system using downscaled data \citep{Berrocal2010AModels}. We then created weekly average exposures for each pregnancy beginning on the date of conception for the census tract of residence. PM$_{2.5}$ exposures were log-transformed to reduce skew. 

We are interested in the exposure-time-response relationship between PM$_{2.5}$ exposure during pregnancy and resulting birth weight. We consider as an outcome birth weight for gestational age $z$-score (BWGAZ) derived using a standard reference table from the gestational age and sex contained in the birth record \citep{Fenton2013AInfants}. We limit our analysis to singleton, full-term births ($\geq37$ weeks) with complete covariate and exposure data. We further restrict our dataset to Northern front range counties (immediately east of the Rocky Mountains roughy extending from Colorado Springs to the Wyoming border), which contains the majority of the Colorado population, and exclude census tracts higher than 6000 feet above sea level, which reduces the potential confounding by altitude and impact of mountainous terrain on exposure predictions. The resulting birth cohort contains 310,236 births---a large dataset by many standards giving ample power for estimating the exposure-time-response relationship between PM$_{2.5}$ and the birth outcome. 

\subsection{A Traditional DLM Analysis}\label{sec:tdlm_analysis}
To complete a standard DLM analysis, without effect heterogeneity, we consider a sample indexed by $i=1,\ldots,n$ with outcome $y_i$, a vector of exposure measurements $\mathbf{x}_i=[x_{i1},\ldots,x_{iT}]'$ taken at equally spaced times $t\in\{1,\ldots,T\}$, and a vector of covariates, $\mathbf{z}_i$, which includes the model intercept. Here, index $i$ refers to a singleton live birth delivery, $y_i$ refers to BWGAZ, while $x_{it}$ is the PM$_{2.5}$ exposure measurement related to observation $i$ during week $t$ of pregnancy. For a Gaussian model, the discrete time DLM takes the form $y_i=\sum_{t=1}^T x_{it}\theta_t+\mathbf{z}_i'\boldsymbol\gamma+\varepsilon_i$, where $\boldsymbol\theta=[\theta_1,\ldots,\theta_T]$ represents the distributed lag function with $\theta_t$ the linear effect due to exposure at time $t$; $\boldsymbol\gamma$ is a vector of regression coefficients; and $\varepsilon_i$ represents independent errors distributed $\mathcal{N}(0,\sigma^2)$. In our analysis, $\mathbf{z}_i$ includes a intercept for county, year, and month of conception, as well as census tract elevation and trimester average temperature. %Additional details on this analysis are given in Section \ref{sec:data_analysis}.

To account for the autocorrelation between exposures measured at higher temporal resolution the distributed lag function, $\boldsymbol\theta$, is typically constrained to vary smoothly over the time period of exposure. These constraints yield effect estimates that are more biologically plausible, adds stability to the estimator in the presence of high autocorrelation in the exposure data, and allows for data-driven identification of critical windows \citep{Wilson2017PotentialHealth}. Methods for constraining DLMs include splines \citep{Zanobetti2000}, Gaussian processes \citep{Warren2020}, principal components \citep{Wilson2017a}, and regression trees \citep{Mork2023EstimatingPairs}. Here, we take the latter approach to DLM as regression trees have been shown to more precisely identify critical windows. %The majority of studies that apply these methods assume a homogeneous exposure-response relationship across the population.

\begin{figure}[!ht]
    \centering
    \includegraphics[height=6cm]{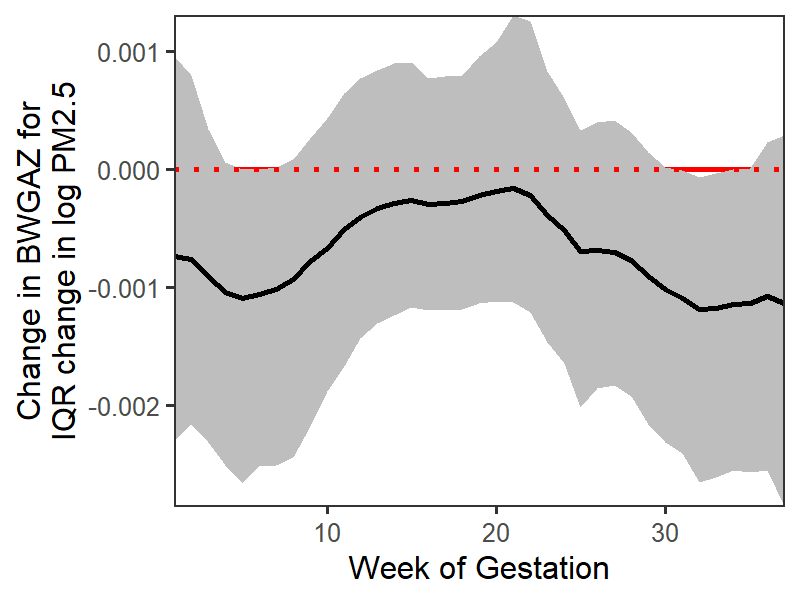}
    \caption{Estimated distributed lag function for the Colorado birth cohort. The solid line indicates the posterior mean while the gray area represents a 95\% credible interval.}
    \label{fig:est_tdlm}
\end{figure}

Based on the DLM analysis without effect heterogeneity, we found increased PM$_{2.5}$ exposure was associated with decreased BWGAZ during each week of gestation. The estimated distributed lag function (Figure \ref{fig:est_tdlm}) identified critical windows during weeks 5-6 and 31-34. For an IQR increase in PM$_{2.5}$ at every week of pregnancy (referred to as the cumulative effect) we see an estimated change in BWGAZ of $-0.026$ (95\% CI: $-0.044,-0.006$). In a more interpretable context based on our observed exposure data, an increase from 5.9 to 8.5 $\mu g/m^3$ PM$_{2.5}$ (the 25$^{th}$ and 75$^{th}$ weekly exposure percentiles, respectively) is related to an approximate decrease in birth weight of 11.3g; this is approximate because BWGAZ adjusts for sex and gestational age.

Existing methods to assess heterogeneity in critical windows rely on estimated subgroup-specific effects using models that are stratified by a single categorical factor or cross-classification based on a small number of factors. Sequential, separate analyses of multiple modifiers have several disadvantages, including issues of multiple testing, the fact that estimates of effect modification by one factor are not controlled by modification by a second correlated modifier, and the risk of failing to identify subgroups that are defined only by combinations of multiple modifying factors. Our dataset is rich with candidate modifiers and we seek to identify vulnerable populations though a coherent analysis that considers multiple candidate modifiers simultaneously. 

%At this point, many studies employing a DLM analysis would stop and be satisfied with the described results, or perhaps proceed to evaluate a few predefined subgroups. However, our dataset is rich in individual-level characteristics and combined with a large sample it allows us to ask more nuanced questions. For example, which individual characteristics modify the PM$_{2.5}$---BWGAZ relationship? Do critical windows differ among susceptible populations? Can we characterize individual-specific exposure-time-response relationships?

\subsection{Heterogeneous Distributed Lag Model}\label{sec:hdlm}
We now describe a heterogeneous distributed lag framework to answer the questions posed in Section \ref{sec:intro}. Consider a set of potential modifiers of the exposure-time-response relationship, denoted $\mathbf{m}_i$ with $\mathbf{m}_i\subseteq \mathbf{z}_i$. These modifiers are used as inputs into the distributed lag function, changing the magnitude and critical windows of the exposure effects. If $\mathbf{m}_i$ is high dimensional, it may also provide unique information about each observation in our analysis or small subgroups with specific characteristics. The heterogeneous distributed lag model (HDLM) is  
\begin{equation}
    \label{eq:hetergeneous_dlm}
    y_i=\sum_{t=1}^T x_{it}\theta_t(\mathbf{m}_i)+\mathbf{z}_i'\boldsymbol\gamma+\varepsilon_i
\end{equation}
where $\boldsymbol\theta(\mathbf{m}_i)=[\theta_1(\mathbf{m}_i),\ldots,\theta_T(\mathbf{m}_i)]$ is a heterogeneous distributed lag function with $\theta_t(\mathbf{m}_i)$ parameterizing the linear effect of exposure at time $t$ specific to an observation with modifiers $\mathbf{m}_i$. Determining when $\theta_t(\mathbf{m}_i)\neq \theta_t(\mathbf{m}_i')$ for two sets of modifiers $\mathbf{m}_i$ and $\mathbf{m}_i'$ will help us to determine differences in the exposure effects and identify vulnerable populations. 

Several methods have sought to estimate DLMs that vary across a population. \cite{Wilson2017a} proposed a Bayesian distributed lag interaction model (BDLIM) to estimate differences in the exposure-response function for a parsimonious set of predetermined subgroups ($\mathbf{m}_i$ represents a single categorical variable). \cite{Warren2020} developed spatially-varying DLMs to account for changes in pollution composition or demographics over a study region ($\mathbf{m}_i$ represents a single categorical variable with spatial correlation matrix for areal data). These papers highlight the bias incurred by a homogeneous effects assumption when effects are truly heterogeneous; however the methods of \cite{Wilson2017a} and \cite{Warren2020} remain limited in their scope and interpretability. A spatially-varying DLM is unable to identify the subject characteristics associated with changes in the underlying distributed lag function. BDLIM is limited to predefined categorical subgroups and most reasonably applied to data having only a small number of subgroups. Currently, there are no approaches to estimate this model within a data-adaptive statistical framework allowing for possibly high dimensional $\mathbf{m}_i$ that may include continuous, categorical, and ordinal candidate modifiers, while also providing necessary constraints for estimation of the distributed lag function.

\section{Regression Tree Approach to HDLM}\label{sec:reg-tree-hdlm}

%We propose a Bayesian additive regression tree (BART) method for estimating distributed lag function heterogeneity due to a set of modifying covariates. 
\subsection{Existing Methods}
\begin{figure}[!ht]
\spacingset{1}
    \centering
    \begin{subfigure}{0.39\textwidth}
        \centering
        \includegraphics[height=6cm]{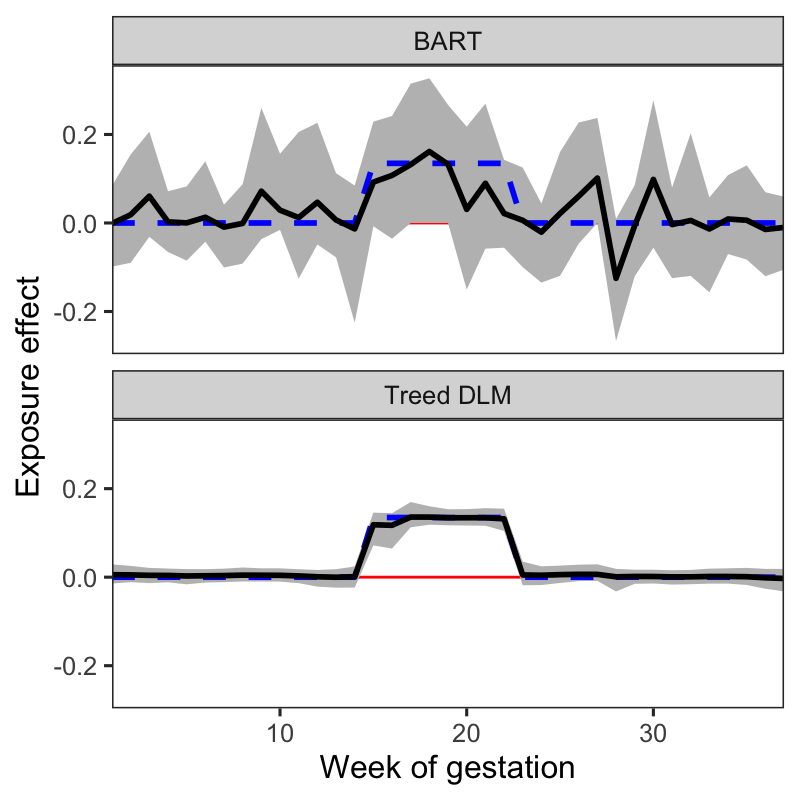}
        \caption{}
        \label{fig:bart_vs_tdlm}
    \end{subfigure}
    \begin{subfigure}{0.59\textwidth}
        \centering
        \includegraphics[height=6cm]{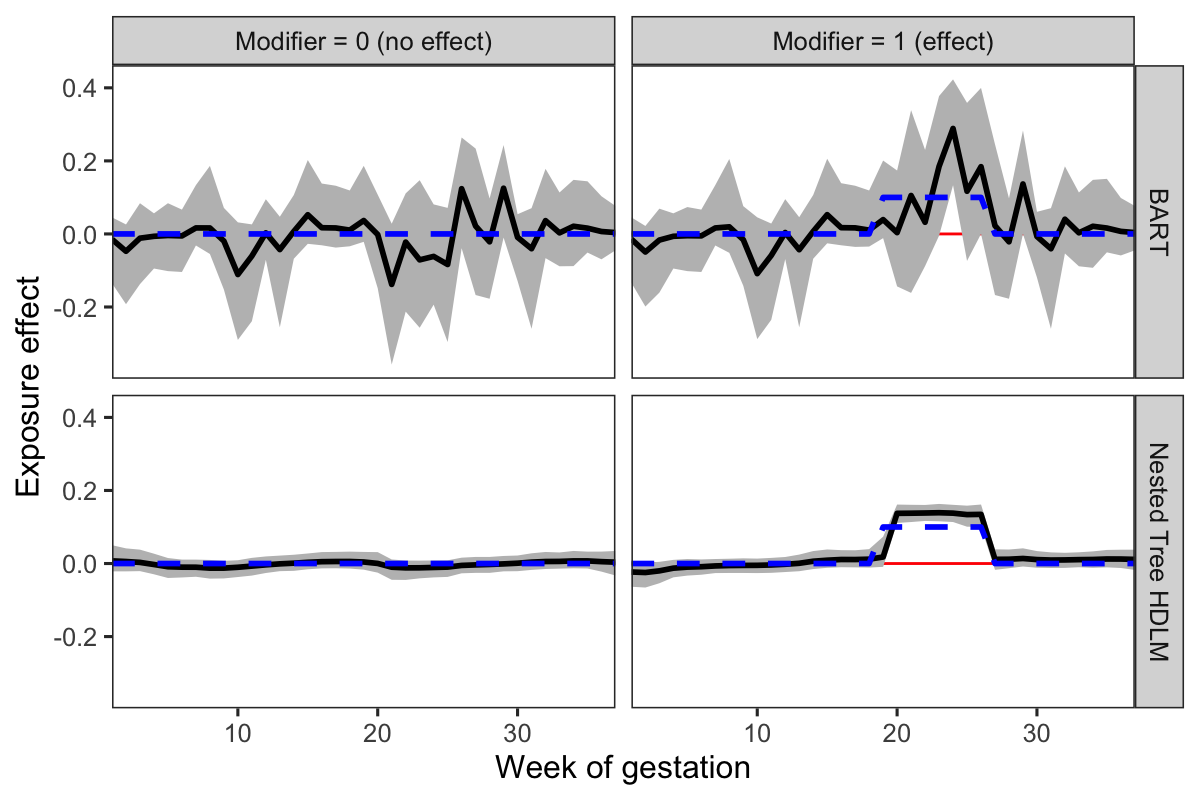}
        \caption{}
        \label{fig:bart_vs_hdlm}
    \end{subfigure}
    \caption{Panel (a) compares estimates from BART with the treed DLM approach of \cite{Mork2023EstimatingPairs} for a distributed lag function without heterogeneity. Panel (b) compares estimates from BART with our approach to heterogeneous DLM when a single binary covariate (columns) modifies the distributed lag function. The dashed blue line displays the true simulated distributed lag function while the solid black line indicates the posterior mean with 95\% credible interval in the shaded region.}
    \label{fig:scen4_5}
\end{figure}

The BART framework of \cite{Chipman2012} is a popular method for estimating non-parametric functions, but applied to our data without any modification is insufficient in several respects. First, BART would treat exposure measurements at separate time points as individual covariates, ignoring the temporal structure of the data and leading to increased variance in the exposure effect estimates. Second, BART does not differentiate between the exposures and modifiers. This results in the same level of regularization being applied to the exposure effect and modification effects and a decreased ability to identify modifying factors. Third, DLMs assume linear effects of exposure at each time point with no interaction between exposures at different time points. In contrast, BART would allow for nonlinear associations and interactions across time, complicating interpretation and inference. \citet{Mork2023EstimatingPairs} and \citet{Mork2022TreedModels} introduced extensions of BART (Treed DLM) to harness the temporal structure of the exposure data, leading to significant improvements over BART as well as other competing distributed lag methods for estimating a distributed lag function (e.g. see Figure \ref{fig:bart_vs_tdlm}). The performance gains that result from adding structure to a BART model for Treed DLM motivate this work by adding structure to a BART model to estimate heterogeneous distributed lag functions.

Several BART methods have been proposed to modify the effect of a single or multiple predictors. \cite{Chipman2002} proposed treed regression models that modify a vector of individual regression coefficients using a single Bayesian tree. Along similar lines, \cite{Deshpande2020VCBART:Coefficients} proposed a varying coefficient BART model that uses a separate ensemble of regression trees to modify each regression coefficient in the model. \cite{Starling2019} introduced BART with targeted smoothing (tsBART), which allows a smooth risk function of a single univariate predictor to vary across a population.  In a non-Bayesian approach to estimate effect heterogeneity, \cite{Odden2020HeterogeneousAdults} applied a random forest algorithm to identify heterogeneous exposure associations. However, these methods cannot account for a structured vector of regression coefficients, such as a constrained distributed lag function, using an ensemble of trees.

\subsection{Bayesian Additive Regression Tree HDLM}\label{sec:bart_hdlm}
Our method for modeling heterogeneity due to a set of modifying covariates uses an ensemble of regression trees combined with functional estimators of the distributed lag effects. We denote a modifier regression tree by $\mathcal{M}_a$ for $a\in\{1,\ldots,A\}$. The modifier regression tree partitions the population based on a set of candidate modifiers $\mathbf{m}_i$. We denote the terminal nodes of $\mathcal{M}_a$ as $\eta_{ab}$ where $b=1,\ldots,B_a$ indexes the subgroups partitioned by tree $a$.  Each terminal node is associated with a $T$-dimensional vector of parameters $\boldsymbol\theta_{ab}$. Considering all trees in the ensemble, the distributed lag function for observation $i$ is
\begin{equation}\label{eq:dlmtree_heterogeneous_dlm}
    \boldsymbol\theta(\mathbf{m}_i)=\sum_{a=1}^A\sum_{b=1}^{B_a}\boldsymbol\theta_{ab}\mathbb{I}(\mathbf{m}_i\in\eta_{ab})
\end{equation}
where $\boldsymbol\theta_{ab}=[\theta_{ab1},\ldots,\theta_{abT}]'$ parameterizes the partial distributed lag function for the subgroup contained in terminal node $\eta_{ab}$ in tree $\mathcal{M}_a$ and $\mathbb{I}(\cdot)$ is the indicator function.

%======================================
\subsection{Nested Tree HDLM}\label{sec:nested_tree_hdlm}
%======================================

\begin{figure}
\spacingset{1}
    \centering
    \begin{subfigure}[b]{.44\textwidth}
    \includegraphics[height=4cm]{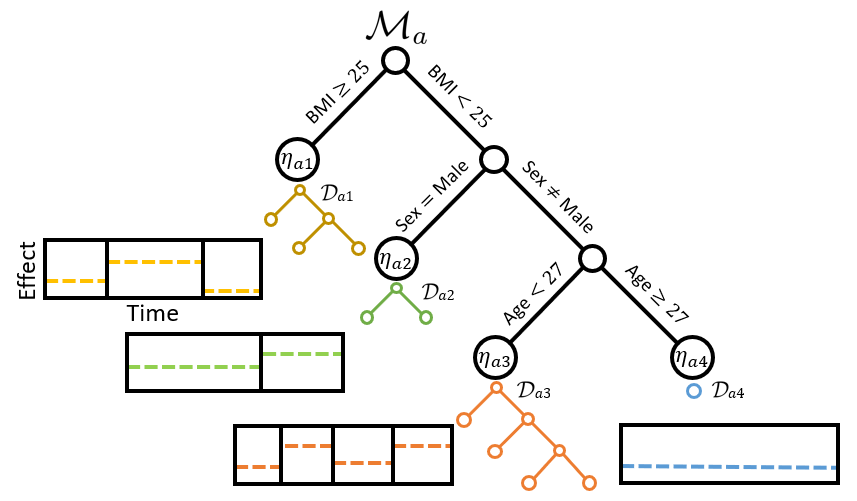}
    \caption{Nested Tree HDLM}
    \label{fig:nested_hdlm}
    \end{subfigure}
    \begin{subfigure}[b]{.54\textwidth}
    \includegraphics[height=4cm]{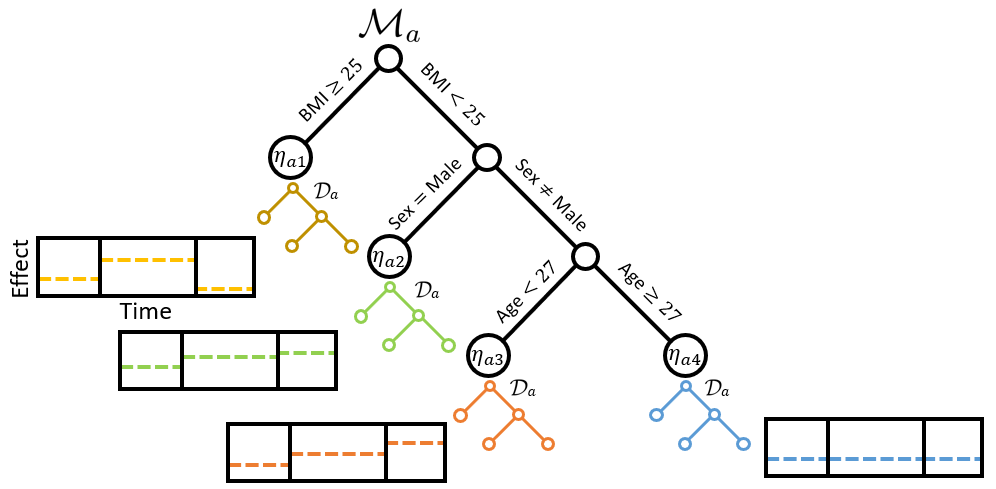}
    \caption{Shared Tree HDLM}
    \label{fig:shared_hdlm}
    \end{subfigure}
    \caption{Panel (a) diagrams the nested tree HDLM. Modifier tree $\mathcal{M}_a$ is structured with binary splitting rules on modifiers BMI, sex, and age. Each terminal node $\eta_{ab}$ has a unique treed DLM structure $\mathcal{D}_{ab}$ with corresponding piecewise effects given by $\delta_{abc}$, shown as dashed lines. Panel (b) diagrams the shared tree HDLM. Here, each terminal node of the modifier tree $\eta_{ab}$ uses the same treed DLM structure $\mathcal{D}_a$. The time points where the distributed lag function changes are shared across all trees while the effect magnitude is unique to each subgroup.}
\end{figure}

\cite{Mork2023EstimatingPairs} showed that treed DLMs outperform competing spline and Gaussian process-based methods in terms of distributed lag function estimation and precision of critical window identification. We propose a treed DLM as a parametric model at each terminal node of the modifier tree. This results in a nested tree structure visualized in Figure \ref{fig:nested_hdlm}. Here, each subgroup of the modifier tree is paired with a unique distributed lag tree structure and corresponding effects.

Before describing the nested tree HDLM we give an overview of the treed DLM. The notation presented here distinguishes the DLM regression tree from the modifier regression tree presented in Section \ref{sec:bart_hdlm}. Consider binary tree $\mathcal{D}$, which partitions the exposure time span, $t=1,\ldots,T$, into $C$ non-overlapping time segments. Internal nodes of $\mathcal{D}$ are assigned binary rules at time points within the period of exposure (e.g. $t<t_1$ and $t\geq t_1$ for $t_1\in\{2,\ldots,T\}$). The terminal nodes of $\mathcal{D}$, denoted $\lambda_c$ for $c\in\{1,\ldots,C\}$, bin together time points to define a piecewise constant distributed lag function (see Figure \ref{fig:nested_hdlm}). That is, $\theta_t=\delta_c$ if $t\in\lambda_c$, where $\delta_c$ is a distributed lag effect for all exposure observations within the time period of $\lambda_c$. Binning of distributed lag effects adds structure to the distributed lag function and stabilizes the model in the presence of autocorrelation in the exposure data.

In the nested tree DLM, each modifier tree subgroup has a unique treed DLM structure. As a result, there is no sharing of information across terminal nodes concerning the timing of the critical window, the effect size, or the smoothness of the DLM. This creates a highly flexible model that allows structures of the treed DLM to adapt to the subgroup whose exposure-response is being estimated. Keeping the number of treed DLM terminal nodes small introduces constraints in the distributed lag function to regularize the model.

To formally define the nested tree DLM in the context of an ensemble of modifier trees, denote $\mathcal{D}_{ab}$ as the treed DLM associated with terminal node $\eta_{ab}$ in modifier tree $\mathcal{M}_a$. Treed DLM $\mathcal{D}_{ab}$ contains $C_{ab}$ terminal nodes, denoted $\lambda_{abc}$, with corresponding distributed lag effects $\delta_{abc}$. To calculate \eqref{eq:dlmtree_heterogeneous_dlm}, let $\theta_{abt}=\delta_{abc}$ if $\mathbf{m}_i\in\eta_{ab}$ and $t\in\lambda_{abc}$. The distributed lag effect for observation $i$ at time $t$ is $\theta_t(\mathbf{m}_i)=\sum_{a=1}^A\sum_{b=1}^{B_a}\sum_{c=1}^{C_{ab}}\delta_{abc}\mathbb{I}(\mathbf{m}_i\in\eta_{ab},t\in\lambda_{abc})$.

%======================================
\subsection{Prior Specification}\label{sec:prior}
%======================================

The nested tree HDLM prior consists of five components: modifier tree structure, treed DLM structures, distributed lag effects, fixed effects of covariates, and error variance. We have several goals when specifying priors. First, as with BART, trees with fewer terminal nodes help to stabilize the model. This is particularly true for treed DLMs where few terminal nodes provide a necessary constraint on the distributed lag function. Second, the model should prioritize rules on modifiers resulting in different DLM effects to remove modifiers that do not differentiate subgroups. Third, we want to lower false window detection by shrinking effects for subgroups and treed DLMs that poorly fit the data.

The priors on modifier tree structures and treed DLM structures follow \citet{Chipman2012}. We apply the Dirichlet prior on splitting rules of \citet{Linero2018b} in the modifier tree to allow for modifier selection. Our model returns both posterior inclusion and variable selection probabilities and in line with \citet{Linero2018b} we make inference using posterior inclusion probabilities. Complete details are given in Supplementary Materials Section 1.

The distributed lag effects are assigned the conjugate normal prior, $\delta_{abc} |\tau_a, \nu, \sigma \sim\mathcal{N}(0,\tau_a^2\nu^2\sigma^2)$,
where $\tau_a,\nu\sim\mathcal{C}^+(0,1)$ define a horseshoe-like estimator on tree specific effects. We include error variance $\sigma^2$ to allow integration over this parameter during tree updates. Our prior specification differs from previous BART implementations \citep{Chipman2012,Starling2019}, which apply uniform variance across all trees. The tree-specific variance prior improves performance in treed DLM by shrinking effects of misspecified trees, reducing variance and false window detection \citep{Mork2023EstimatingPairs}. 

To complete a fully Bayesian specification of the nested tree DLM we assign a non-informative prior to the fixed effects, $\boldsymbol\gamma\sim\mathcal{MVN}(\mathbf{0},d\sigma^2 I_p)$, where $I_p$ is a $p\times p$ identity matrix and $d$ is fixed at a large value. Finally, we specify prior $\sigma\sim\mathcal{C}^+(0,1)$. Complete details are given in Supplementary Materials Section 1.

%======================================
\subsection{Computation}
%======================================

We estimate nested tree HDLM by sampling from the posterior distribution using Markov chain Monte Carlo (MCMC) methods. As in BART, we apply Bayesian backfitting \citep{Hastie2000} to estimate effects for each modifier tree and apply an independent Metropolis-Hastings (MH) algorithm to update modifier and treed DLM structures. Our algorithm differs from other BART implementations in several ways. First, we control for a set of fixed effects when estimating the heterogeneous DLMs. Second, each terminal node of the modifier tree has a unique treed DLM structure that is learned from the data and must be updated. We briefly outline our algorithm and provide full details in Supplementary Materials Section 2.

To account for potential changes in dimensionality during modifier tree updates we integrate over all terminal node effects, $\delta_{abc}$. The update can then proceed with a MH step similar to that used in BART. To improve updates for tree structures and estimates of distributed lag functions we integrate the fixed effect parameters, $\boldsymbol\gamma$, and model variance, $\sigma^2$, out of the data likelihood. Updates to modifier tree structures occur through the four proposal steps described in \cite{Chipman2012}: grow, prune, change, and swap. The grow step adds an additional split at a terminal node; prune removes a split from an internal node connected to two terminal nodes; change step modifies a binary splitting rule at an internal node; and swap reverses the order of rules in two adjacent internal nodes. 

In the nested tree HDLM, each terminal node of the modifier tree has a unique treed DLM structure. In a grow proposal, a terminal node becomes an internal node and the corresponding treed DLM structure is eliminated and replaced by two new treed DLM structures at the new modifier tree terminal nodes. For each new terminal node, a new treed DLM structure is drawn from the tree prior. Conversely for a prune proposal, an internal node becomes a terminal node. In this case the two existing treed DLM structures are discarded and a new treed DLM structure is drawn from the tree prior. During a change or swap step in the modifier tree the terminal nodes retain the same treed DLM structure. 

After updating a modifier tree, we update treed DLM structures associated with each modifier tree terminal node. Updates to treed DLM structures use grow, prune, and change proposals because swap results in empty terminal nodes. Conditionally conjugate normal priors allow for multivariate Gibbs update of the distributed lag functions. Following updates to tree structures and terminal node effects, the remaining parameters are updated with standard MCMC procedures. Convergence is assessed by comparing results of multiple MCMC chains and examining trace plots of distributed lag estimates, see Supplemental Materials Section 4 for an example from our case study.

%======================================
\subsection{Shared Tree HDLM}\label{sec:shared_tree_hdlm}
%======================================

The shared tree HDLM is a simplification of the nested tree HDLM, described in Supplementary Materials Section 1, that uses the same treed DLM structure for each terminal node of a modifier tree but the distributed lag coefficients are allowed to be different magnitudes. The shared tree HDLM is visualized in Figure \ref{fig:shared_hdlm}. As in the nested tree HDLM, tree structures are learned from the data, but a change in the treed DLM structure is applied to all terminal nodes of a modifier tree. This assumes that the distributed lag functions have the same change points for critical windows. For our data problem involving prenatal development, this could relate to an assumption of developmental stages always occurring during the same weeks, although the magnitude of effects may differ. In contrast, nested tree HDLM allows for the possibility that onset and duration of the effects varies by subgroup.

%======================================
\subsection{Alternative Parameterizations}\label{sec:gp_hdlm}
%======================================
We explored alternative parameterizations to incorporate a constrained DLM in our ensemble model, including Gaussian process and regression spline approaches. Regression splines have a tendency misclassify critical windows and were not considered for this application. A Gaussian process HDLM showed some promise but was consistently subpar to the nested and shared tree HDLM. Complete details of the Gaussian process alternative approach are given in Supplementary Materials Section 1.

%An alternative to treed DLMs is a Gaussian process approach to HDLM. Consider a single modifier tree terminal node, $\eta_{ab}$, containing a subset of observations. Let the corresponding $T-$dimensional set of DLM parameters $\boldsymbol\theta_{ab}$ have Gaussian process prior, $\boldsymbol\theta_{ab} |\tau_a,\nu,\sigma,\phi \sim \mathcal{GP}[\mathbf{0},\tau_a^2\nu^2\sigma^2\boldsymbol\Sigma(\phi)]$, where $\Sigma_{\boldsymbol\phi}(t,t')$, is a covariance function with parameters $\phi$, and $\tau_a,\nu$ are horseshoe-like estimators as described in Section \ref{sec:prior}. An inherent assumption of the Gaussian process method is that all observations share the same value of $\phi$ due to the computation approach and therefore have the same smoothness over time in their distributed lag functions. The equal smoothness assumption may be beneficial if the distributed lag function for all observations have a similar magnitude and window length. In the case where there is a nonzero exposure effect in only a small proportion of the population, the smoothness of the distributed lag function imposed by the remaining sample will make this critical window difficult to estimate. This is because estimation of the smoothing parameter will largely reflect the null group and cause over-smoothing in the group for which there is a non-zero association during some gestational weeks. Complete details of the Gaussian process method are given in Supplementary Materials Section 1.

%======================================
\section{Simulation Studies}\label{sec:sim}
%======================================

We developed five simulation scenarios to evaluate estimation of the heterogeneous distributed lag function and the ability to identify correct modifiers. The first simulation scenario relates to the nested tree HDLM and considered subgroups with three different distributed lag functions: an early window, a late window, and no effect. The second scenario imitated shared tree HDLM and had two groups: a distributed lag function with a critical window that does not change in time but was scaled by a continuous modifier and a group with no effect. Scenario 3 compared HDLM to traditional DLM methods when there was no effect modification. Scenarios 4 and 5 compare HDLM and standard treed DLM to BART when estimating the distributed lag function with or without heterogeneity. Details regarding scenarios 3, 4 and 5 are presented in Supplementary Materials Section 3.

In general, we found that nested and shared tree HDLMs performed similarly to a standard DLM in scenario 3 with no effect modification. In a low noise setting we saw a greater distinction between the nested and shared tree approaches, while they were comparable in a high noise scenario. In scenarios 4 and 5 we found the tree-based DLMs identified critical windows and between group differences with high probability whereas BART often failed to identify windows or between group differences. Figures \ref{fig:bart_vs_tdlm} and \ref{fig:bart_vs_hdlm} show a single simulation replicate from scenarios 4 and 5, respectively. Full simulation results from scenarios 4 and 5 are given in Supplementary Materials Section 3.

Scenarios 1, 2, and 3 involved 13 covariates. Two of these covariates were responsible for the DLM heterogeneity in scenarios 1 and 2. Covariates $\mathbf{z}_i=[z_{i1},\ldots,z_{i13}]'$ were independently generated: $z_{i1}$ $\sim\mathcal{N}(0,1)$, $z_{i2}$ $\sim\text{Bernoulli}(0.5)$, $z_{i3}$ $\sim\text{Uniform}(0,1)$, $z_{ip}\sim\mathcal{N}(0,1)$ for $p\in\{4,\ldots,8\}$, $z_{ip}\sim\text{Bernoulli}(0.5)$ for $p\in\{9,\ldots,13\}$. We also include $z_{i0}=1$ as a model intercept. A set of 37 consecutive weekly PM$_{2.5}$ exposures, denoted $\mathbf{x}_i=[x_{i1},\ldots,x_{i37}]'$ was drawn for each observation using real exposure measurements from our data analysis.

The simulation generated continuous outcomes under the model $y_i= r\cdot\mathbf{x}_i'\boldsymbol\theta(\mathbf{m}_i)+\mathbf{z}_i'\boldsymbol{\gamma}+\varepsilon_i$,
where $\boldsymbol{\gamma}$ are parameters drawn from standard normal, and $\boldsymbol\theta(\mathbf{m}_i)$ represent the HDLM from each simulation scenario. In the HDLM we let $\mathbf{m}_i$ be all variables in $\mathbf{z}_i$ except for the intercept, although only two modifiers were actually responsible for the effect heterogeneity. Scaling factor $r$ was defined such that Var$[r\cdot \mathbf{x}_i'\boldsymbol\theta(\mathbf{m}_i)]=1$, and $\varepsilon_i$ was drawn independently from $\mathcal{N}(0,\sigma^2)$ under two different noise settings, $\sigma^2\in\{1,25\}$. Each simulation scenario and $\sigma^2$ combination was run for 100 replicates with $n=$ 5,000 observations.

We applied three models to each simulated dataset: nested tree HDLM, shared tree HDLM and a standard treed DLM for no effect heterogeneity. In Supplementary Materials Section 3 we show extended results including Gaussian process HDLM and DLM. For each method described, we considered two model selection criteria: mean square prediction error (MSPE) and Watanabe-Akaike information criterion (WAIC) \citep{Watanabe2010AsymptoticTheory}. MSPE was calculated as $n^{-1}\sum_{i=1}^n(y_i-\hat{y}_i)^2$, using 5,000 out-of-sample observations while WAIC was calculated on the same 5,000 observations used for model fitting. Simulation scenario 1 was also estimated using the nested tree HDLM where the modifier tree was fixed to use the true subgroups. All models used 20 modifier trees in the ensemble and were run for 10,000 MCMC iterations thinned to every 5$^{th}$ iteration, following 5,000 burn-in iterations. Increasing the number of trees did not improve performance. All simulations can be reproduced with \texttt{R} package \texttt{dlmtree}.

%======================================
\subsection{Scenario 1: Early/Late Window}
%======================================

We simulated the heterogeneous distributed lag function with two true modifiers,  $z_{i1}$ and $z_{i2}$, as $\theta_t(\mathbf{m}_i)= \mathbb{I}(t\in[11,18])\text{ if }z_{i1}>0\text{ and }z_{i2}=1$; $\theta_t(\mathbf{m}_i)= \mathbb{I}(t\in[17,26])\text{ if }z_{i1}>0\text{ and }z_{i2}= 0$; and $\theta_t(\mathbf{m}_i)= 0\text{ if } z_{i1}\leq 0$.
The first group ($z_{i1}>0, z_{i2}=1$) had a nonzero distributed lag function during weeks 11--18. The second group ($z_{i1}>0, z_{i2}=0$) had a nonzero distributed lag function during weeks 17--26, overlapping with the first group by two weeks. The third group ($z_{i1}\leq 0$) had no exposure effect.

Pointwise DLM results averaged across simulation replicates are given in Table \ref{tab:scen1_res}. We separately analyzed subgroups with an effect ($z_{i1}>0$) from subgroups with no effect ($z_{i1}\leq0$). We report DLM root mean square error (RMSE) $=\sqrt{\sum_{t=1}^{37}[\theta_t(\mathbf{m}_i)-\hat{\theta}_t(\mathbf{m}_i)]^2/37}$ and coverage of the distributed lag function by 95\% pointwise credible intervals, averaged across observations in effect and no effect subgroups. We also calculated the probability that the model detects a true critical window (true positive, TP) when $\theta_t(\mathbf{m}_i)$ is nonzero as well as the probability a critical window is identified where the true effect is zero (false positive, FP), using the 95\% pointwise credible interval for the distributed lag function. To gauge effectiveness of MSPE and WAIC for model selection we report proportion of replicates where each model is selected and summarize results for the selected model.

\begin{table}[!ht]
\spacingset{1}
\scriptsize
    \centering
    \caption{Simulation results for estimating the DLM in scenario 1 (early/late effect). Results are considered pointwise across the DLM for each individual and broken down for individuals with a zero versus nonzero effect. Selected indicates the proportion of replicates for which a given method was selected via MSPE or WAIC.}\vspace{6pt}
    \label{tab:scen1_res}
    \begin{tabular}{lrrrrrrrrrcc}
        \toprule[2pt]
        &\multicolumn{4}{c}{Effect ($z_{i1}>0$)}&&\multicolumn{3}{c}{No Effect ($z_{i1}\leq0$)}&&\multicolumn{2}{c}{Selected}\\
        \cmidrule(lr){2-5} \cmidrule(lr){7-9} \cmidrule(lr){11-12}
        Model & RMSE$^*$ & Coverage & TP & FP & \phantom{} &RMSE$^*$ & Coverage & FP && MSPE & WAIC\\
        \midrule
        \multicolumn{9}{l}{$\sigma^2=1$}\\
      Nested Tree HDLM & 1.94 & 0.97 & 1.00 & 0.02 &  & 1.88 & 0.99 & 0.01 &  & 0.65 & 0.53\\
      Shared Tree HDLM & 2.15 & 0.97 & 0.99 & 0.02 &  & 1.97 & 0.97 & 0.03 &  & 0.35 & 0.47\\
 %Gaussian Process HDLM & 3.24 & 0.95 & 0.99 & 0.03 &  & 2.46 & 0.98 & 0.02 &  & 0.01 & 0.00\\
%\addlinespace
            Treed DLM & 10.87 & 0.61 & 1.00 & 0.22 &  & 5.71 & 0.61 & 0.39 &  & 0.00 & 0.00\\
               %Gaussian Process DLM & 10.97 & 0.61 & 1.00 & 0.23 &  & 5.67 & 0.61 & 0.39 &  & 0.00 & 0.00\\
\addlinespace
MSPE Selected Model & 2.00 & 0.97 & 1.00 & 0.02 &  & 1.86 & 0.98 & 0.02 &  \\
WAIC Selected Model & 2.05 & 0.97 & 0.99 & 0.02 &  & 1.93 & 0.98 & 0.02 &  \\
\addlinespace
   Nested Tree: Truth & 0.74 & 0.99 & 1.00 & 0.01 &  & 0.47 & 1.00 & 0.00 &  \\
     %Gaussian Process: Truth & 2.76 & 0.96 & 1.00 & 0.03 &  & 1.77 & 0.98 & 0.02 &  \\

\midrule
        
        \multicolumn{9}{l}{$\sigma^2=25$}\\
        Nested Tree HDLM & 6.29 & 0.92 & 0.90 & 0.04 &  & 2.67 & 1.00 & 0.00 &  & 0.75 & 0.77\\
        Shared Tree HDLM & 7.03 & 0.91 & 0.88 & 0.05 &  & 3.25 & 0.99 & 0.01 &  & 0.25 & 0.23\\
   %Gaussian Process HDLM & 6.99 & 0.95 & 0.92 & 0.02 &  & 4.02 & 1.00 & 0.00 &  & 0.16 & 0.07\\
%\addlinespace
              Treed DLM & 11.49 & 0.65 & 0.81 & 0.17 &  & 5.42 & 0.69 & 0.31 &  & 0.00 & 0.00\\
   %Gaussian Process DLM & 11.52 & 0.70 & 0.63 & 0.11 &  & 5.47 & 0.78 & 0.22 &  & 0.00 & 0.00\\
\addlinespace
MSPE Selected Model & 6.33 & 0.92 & 0.90 & 0.04 &  & 2.78 & 1.00 & 0.00 &   \\
WAIC Selected Model & 6.27 & 0.92 & 0.90 & 0.04 &  & 2.80 & 1.00 & 0.00 &  \\
\addlinespace
      Nested Tree: Truth & 5.26 & 0.96 & 0.95 & 0.02 &  & 2.10 & 1.00 & 0.00 & \\
 %Gaussian Process: Truth & 6.37 & 0.96 & 0.98 & 0.02 &  & 3.71 & 1.00 & 0.00 & \\
        \bottomrule[2pt]
        \multicolumn{9}{l}{*RMSE$\times100$}\\
    \end{tabular}
\end{table}

The nested tree HDLM yielded distributed lag estimates with smaller RMSE than those from shared tree HDLM across both error settings. For critical window identification, the nested tree HDLM yielded a similar or higher TP than the other models in both error settings. The FP of the nested tree HDLM ranged from 0.02 in the low error setting to 0.04 in the high error setting; the shared tree HDLM had FP ranging from 0.02 to 0.05, respectively. The added flexibility of the nested tree HDLM allowed for different change points for critical windows and varying smoothness in the effect and no effect groups. Coverage of the distributed lag function was near nominal levels although slightly lower in the high error setting. The decreased coverage was only evident in the subgroups with a true exposure effect and is due to two factors: shrinkage of the exposure effect and in some cases combining the two groups with different critical windows into a single group.  Coverage in the no effect group was above the nominal level. 

The nested tree HDLM was selected most often by MSPE and WAIC across both error settings. The non-heterogeneous DLMs were never selected, indicating HDLMs better predict the outcome when heterogeneity exists. The MSPE and WAIC selected models had RMSE, coverage, TP and FP comparable to the best performing model (nested tree HDLM), giving evidence that these are valid tools for model selection. The nested tree HDLM with subgroups fixed at the truth outperformed all other models. %Notably, in the high error scenario, the MSPE selected model on average outperformed the individual models in terms of effect RMSE, coverage, and FP.

%The treed DLM model with subgroups fixed at the truth outperformed the HDLM models. This is first due to the fact that the true distributed lag function is not smooth. Second, the smoothness assumption of the Gaussian process was homogeneous across all subgroups, leading to over-smoothing in the effect subgroups and under-smoothing in the no effect subgroup. These results demonstrate the advantage of treed DLM approaches when considering heterogeneity. The DLM methods with no effect modification were consistently the worst performing models with low coverage of the distributed lag function, higher FP rates, and the highest RMSE.

In Supplementary Materials Section 3 we present modifier posterior inclusion probabilities (PIP) for an individual modifier or interactions of modifiers. The PIP for an individual modifier is the probability that the modifier is used in at least one splitting rule across the ensemble of trees. We define a modifier interaction to be when two modifiers are used as consecutive splitting rules in the same tree. The average modifier PIPs is a function of the number of modifiers and number of trees in the model, with larger PIPs for fewer modifiers or more trees. Modifiers with a larger PIP relative to other modifiers represent possible modification of the distributed lag function and critical windows, and give a starting point for comparing the exposure effects for individuals or subgroups.

In scenario 1, the true modifiers ($z_1$ and $z_2$) had PIPs ranging from 0.97 to 1 across error settings. The other modifiers had PIPs ranging from 0.51 to 0.62, on average. The modifier that determined the critical window placement, $z_2$ had slightly lower PIP than the modifier that determined effect versus no effect groups, $z_1$. For scenario 1, we found an interaction between modifiers $z_1$ and $z_2$ had a PIP of 1 and 0.88 in the low and high error settings, respectively. The average PIP for other interactions ranged from 0.10 to 0.12.

%======================================
\subsection{Scenario 2: Scaled Effect}
%======================================

In scenario 2 we simulated the heterogeneous distributed lag function based on two continuous covariates $z_{i1}$ and $z_{i3}$, as $\theta_t(\mathbf{m}_i)= z_{i3}\mathbb{I}(t\in[11,18])\text{ if }z_{i1}>0$; and $\theta_t(\mathbf{m}_i)= 0\text{ if }z_{i1}\leq 0$. The distributed lag function is nonzero during weeks 11-18 for the first group ($z_{i1}>0$) and scaled by the modifier $z_{i3}$. The second group ($z_{i1}\leq 0$) had no exposure effect. Here, we did not compare to a fixed subgroups model because the continuous modification does not allow for true subgroups.

Table \ref{tab:scen2_res} summarizes model performance in term of estimation of the distributed lag function and model selection by MSPE and WAIC. The shared tree HDLM yielded lowest RMSE on the distributed lag function and highest TP rate for identifying windows among HDLMs. The added flexibility of the nested tree model was not needed, which leads to slightly lower performance of this approach in this scenario. The shared tree HDLM is most often selected by MSPE followed by the nested tree HDLM, which is also a correctly specified model in this scenario. The non-heterogeneous DLM is never selected. We highlight the fact that the MSPE selected model often outperforms the shared tree HDLM in terms of RMSE and coverage, while having similar TP and FP to the shared tree HDLM. These results give further indication that MSPE can be used as a valid model selection tool.

\begin{table}[!ht]
\spacingset{1}
 \scriptsize
    \centering
    \caption{Simulation results for estimating the DLM in scenario 2 (scaled effect). Results are considered pointwise across the DLM for each individual and broken down for individuals with a zero versus nonzero effect. Selected indicates the proportion of replicates for which a given method was selected via MSPE or WAIC.}\vspace{6pt}
    \label{tab:scen2_res}
    \begin{tabular}{lrrrrrrrrrcc}
        \toprule[2pt]
        &\multicolumn{4}{c}{Effect ($z_{i1}>0$)}&&\multicolumn{3}{c}{No Effect ($z_{i1}\leq0$)}&&\multicolumn{2}{c}{Selected}\\
        \cmidrule(lr){2-5} \cmidrule(lr){7-9} \cmidrule(lr){11-12}
        Model & RMSE$^*$ & Coverage & TP & FP & \phantom{a} &RMSE$^*$ & Coverage & FP && MSPE & WAIC\\
        \midrule
        \multicolumn{9}{l}{$\sigma^2=1$}\\
      Nested Tree HDLM & 2.73 & 0.91 & 0.93 & 0.02 &  & 1.73 & 0.99 & 0.01 &  & 0.31 & 0.38\\
      Shared Tree HDLM & 2.50 & 0.92 & 0.95 & 0.02 &  & 1.77 & 0.98 & 0.02 &  & 0.69 & 0.62\\
 %Gaussian Process HDLM & 3.73 & 0.95 & 0.86 & 0.02 &  & 2.40 & 0.99 & 0.01 &  & 0.01 & 0.00\\
%\addlinespace
             Treed DLM & 9.84 & 0.79 & 1.00 & 0.01 &  & 6.39 & 0.78 & 0.22 &  & 0.00 & 0.00\\
               %Gaussian Process DLM & 10.13 & 0.79 & 1.00 & 0.02 &  & 6.25 & 0.76 & 0.24 &  & 0.00 & 0.00\\
\addlinespace
MSPE Selected Model & 2.52 & 0.92 & 0.94 & 0.02 &  & 1.73 & 0.99 & 0.01 &  \\
WAIC Selected Model & 2.62 & 0.91 & 0.95 & 0.02 &  & 1.75 & 0.99 & 0.01 &  \\
        \midrule

        \multicolumn{9}{l}{$\sigma^2=25$}\\
              Nested Tree HDLM & 6.43 & 0.92 & 0.66 & 0.01 &  & 2.60 & 1.00 & 0.00 &  & 0.40 & 0.41\\
      Shared Tree HDLM & 6.23 & 0.93 & 0.73 & 0.01 &  & 2.85 & 0.99 & 0.01 &  & 0.60 & 0.59\\
 %Gaussian Process HDLM & 7.43 & 0.94 & 0.64 & 0.01 &  & 3.90 & 1.00 & 0.00 &  & 0.10 & 0.02\\
%\addlinespace
            Treed DLM & 10.83 & 0.82 & 0.92 & 0.02 &  & 5.81 & 0.79 & 0.21 &  & 0.00 & 0.00\\
               %Gaussian Process DLM & 11.06 & 0.84 & 0.89 & 0.01 &  & 5.76 & 0.80 & 0.20 &  & 0.00 & 0.00\\
\addlinespace
MSPE Selected Model & 6.05 & 0.93 & 0.70 & 0.01 &  & 2.74 & 1.00 & 0.00 &  \\
WAIC Selected Model & 6.32 & 0.92 & 0.72 & 0.01 &  & 2.75 & 0.99 & 0.01 &  \\

        \bottomrule[2pt]
        \multicolumn{9}{l}{*RMSE$\times100$}\\
    \end{tabular}
\end{table}

Supplementary Materials Section 3 presents PIPs for individual modifiers and modifier interactions. The modifier tree HDLMs correctly distinguish the true modifiers in both error settings with PIPs ranging from 0.98 to 1. The other modifiers have PIPs ranging from 0.43 to 0.62, on average. We found the interaction between modifiers $z_1$ and $z_3$ to have PIP of 1 in the low error setting and 0.91 in the high error setting. The average PIP for other modifier interactions ranged from 0.08 to 0.11.

%======================================
\section{Heterogeneous Critical Windows for PM$_{2.5}$ and Birth Weight}\label{sec:data_analysis}

Using the dataset described in Section \ref{sec:co_birth_data} we applied the nested tree and shared tree HDLMs as well as standard treed DLM (as described in Section \ref{sec:tdlm_analysis}) to estimate the relationship between BWGAZ and a mother's exposure to PM$_{2.5}$ during the first 37 weeks of pregnancy. We allowed for effect heterogeneity due to ten modifiers including continuous variables: maternal age and BMI; ordinal variables: income classification, highest educational attainment, and smoking (never, former, less than 10 cigarettes/day, at least 10 cigarettes/day); nominal variables: marital status, prenatal care, and race; and binary variables: sex and Hispanic indicators. We controlled for demographic, spatial, and temporal covariates outlined in Table \ref{tab:data_cov_mod}. We did not include a fixed effect for fetal sex as the outcome, BWGAZ, was already adjusted for this factor. Each model ran for 15,000 iterations after 5,000 burn-in and was thinned to every 5$^{th}$ iteration. Following 10-fold cross-validation, the shared tree HDLM had lowest average MSPE followed by the nested tree HDLM while standard treed DLM had the largest MSPE. Based on our simulation results showing MSPE as a reliable model selection criteria, we report results from shared tree HDLM. Details of our cross-validation and results from nested tree HDLM are provided in Supplementary Materials Section 4.

\begin{table}[!ht]
\spacingset{1}
 \scriptsize
    \centering
    \caption{Covariates included as fixed effects or modifiers (indicated by checks) of the HDLM. For covariates included as modifiers we report the posterior inclusion probability (PIP), which is defined as the probability the modifier is used in at least one splitting rule in the ensemble.}\vspace{6pt}
    \label{tab:data_cov_mod}
    \begin{tabular}{llrcccr}
        \toprule[2pt]
        Covariate & Type & Mean (IQR) & Categories & Fixed effect & Modifier & PIP\\
        \hline
        Age at conception & Continuous & 28.7 ($24-33$) && \checkmark& \checkmark & 0.93\\
        Height & Continuous & 64.4 ($62-66$) && \checkmark& & $-$\\
        Prior weight & Continuous & 151.3 ($126-169$) && \checkmark&  & $-$\\
        Body mass index & Continuous & 25.7 ($21.6-28.4$) && \checkmark& \checkmark & 0.95\\
        Avg. temp/trimester & Continuous & 51.6 ($38.2-65.1$)& & \checkmark& &$-$ \\
        Income range & Ordinal && 6 & \checkmark& \checkmark & 0.74\\
        Highest education & Ordinal && 5 & \checkmark& \checkmark & 0.90\\
        Smoking habits & Ordinal && 4 & \checkmark& \checkmark & 0.78\\
        Marital status & Categorical && 6 & \checkmark& \checkmark & 0.50\\
        Prenatal care & Categorical && 3 & \checkmark& \checkmark & 0.48\\
        Race & Categorical && 4 & \checkmark& \checkmark & 0.61\\
        County of residence & Categorical && 12 & \checkmark& &$-$ \\
        Month of conception & Categorical && 12 & \checkmark& & $-$\\
        Year of conception &  Categorical && 9 & \checkmark& & $-$\\
        Hispanic & Binary && 2 & \checkmark& \checkmark & 0.95\\
        Sex of child & Binary && 2 & &\checkmark & 0.64\\
        \bottomrule[2pt]
    \end{tabular}
\end{table}

%======================================
%\subsection{DLM without effect modification}\label{sec:data_tdlm}
%======================================
%Figure \ref{fig:est_tdlm} shows the estimated exposure effect using a treed DLM with no effect modification. 
%The treed DLM with no effect modification found that increased PM$_{2.5}$ exposure was associated with decreased BWGAZ during each week of gestation. The DLM (Supplemental Figure 7) identified critical windows during weeks 5-6 and 31-34. The cumulative effect of an inter-quartile range (IQR) increase in PM$_{2.5}$ at every week of pregnancy corresponds to a decrease in BWGAZ of $-0.026$ (95\% CI: $-0.044,-0.006$). In a more interpretable context, an increase from 5.9 to 8.5 $\mu g/m^3$ PM$_{2.5}$ (the 25$^{th}$ and 75$^{th}$ weekly exposure percentiles, respectively) relates to an approximate decrease in birth weight of 11.3g; this is approximate because BWGAZ adjusts for sex and gestational age.

%\begin{figure}[!ht]
%\spacingset{1}
%    \centering
%    \includegraphics[height=5cm]{img/bwgaz_tdlm.png}
%    \caption{Estimated distributed lag functions due to an IQR increase in PM$_{2.5}$ using the treed DLM with no effect modification. The solid line indicates the posterior mean while the gray area represents a 95\% credible interval. Weeks where the credible interval does not contain zero represent critical windows.}
%    \label{fig:est_tdlm}
%\end{figure}

%======================================
\subsection{Modifier selection to determine vulnerable populations}\label{sec:data_mod_sel}
%======================================

The modifier PIPs from the shared tree HDLM are presented in Table \ref{tab:data_cov_mod}. The modifiers with the highest PIP include maternal BMI (0.95), Hispanic (0.95), age (0.93) and education (0.90). The next highest PIP modifiers were smoking (0.78) and income (0.74). Considering empirical evidence from our simulation scenarios, with a similar number of modifiers and trees, PIPs below 0.7 for individual modifiers do not provide strong evidence of effect modification. For continuous modifiers, the posterior distribution of splitting rules for age and BMI had modes at 27 and 22.8, respectively. Splitting rules with a higher posterior probability may indicate a larger distinction between two groups at that level and may hold clinical significance allowing for targeted public health awareness. For modifier interactions, our simulation indicated interaction PIPs above 0.5 are relevant to the study. We found the highest interaction PIPs to be education--BMI (0.65), education--age (0.64), and BMI--Hispanic (0.57). Additional details are given in Supplementary Materials Section 4.

When we consider effect modification, we can interpret the effect heterogeneity as an interaction between the modifiers and the exposure effect. For instance, a subgroup that is the result of a rule on a single modifier (59\% of tree-specified subgroups) is a two-way interaction between that modifier and PM$_{2.5}$. The distributed lag function for a subgroup that is the result of rules on two modifiers (33\% of tree-specified subgroups) is a three-way interaction between the two modifiers as well as exposure to PM$_{2.5}$. While the ensemble of trees should be able to incorporate interactions via the additive nature of multiple trees, some complex interactions, such as the effect/no effect partition in our simulations, may only be captured with a single tree that splits on multiple modifiers.

The modifier PIPs provide a starting point to explore potential vulnerable populations, which are characterized by their differential effects to exposure. In this analysis, we focus on four modifiers: maternal BMI, Hispanic, age, and education, along with interactions between these modifiers. To simplify visualization of the results, we divided continuous modifiers, BMI and age, at the splitting value modes. We present subgroup specific average effects along with personalized distributed lag function estimates for a sample of individuals in each subgroup to give a sense of the remaining within-group heterogeneity.

%======================================
\subsection{Subgroup-specific distributed lag functions}
\label{sec:subgroup-analysis}
%======================================
The HDLM framework allows for inference on personalized distributed lag functions for a specific level of modifiers or subgroup specific distributed lag estimates averaged over the levels of modifiers within a particular subgroup. Let $S$ be a set of observations based on a subgroup of interest, e.g. all babies born to Hispanic mothers or all boy babies with obese mothers. Denote $S_{\eta_{ab}}\subset S$ as the observations from $S$ contained in modifier tree terminal node $\eta_{ab}$. We define weights based on the proportion of $S$ in each terminal node, $w(S,\eta_{ab})=|S_{\eta_{ab}}|\big/|S|$. The subgroup average DLM for $S$ is then calculated $\boldsymbol\theta_S=\sum_{a=1}^A\sum_{b=1}^{B_a}\boldsymbol\theta_{ab}w(S,\eta_{ab})$, where $\boldsymbol\theta_{ab}$ is the vector of DLM effects for modifier tree terminal node $\eta_{ab}$.

\begin{figure}[!ht]
\spacingset{1}
    \centering
    \begin{subfigure}{0.49\textwidth}
        \includegraphics[height=5.75cm]{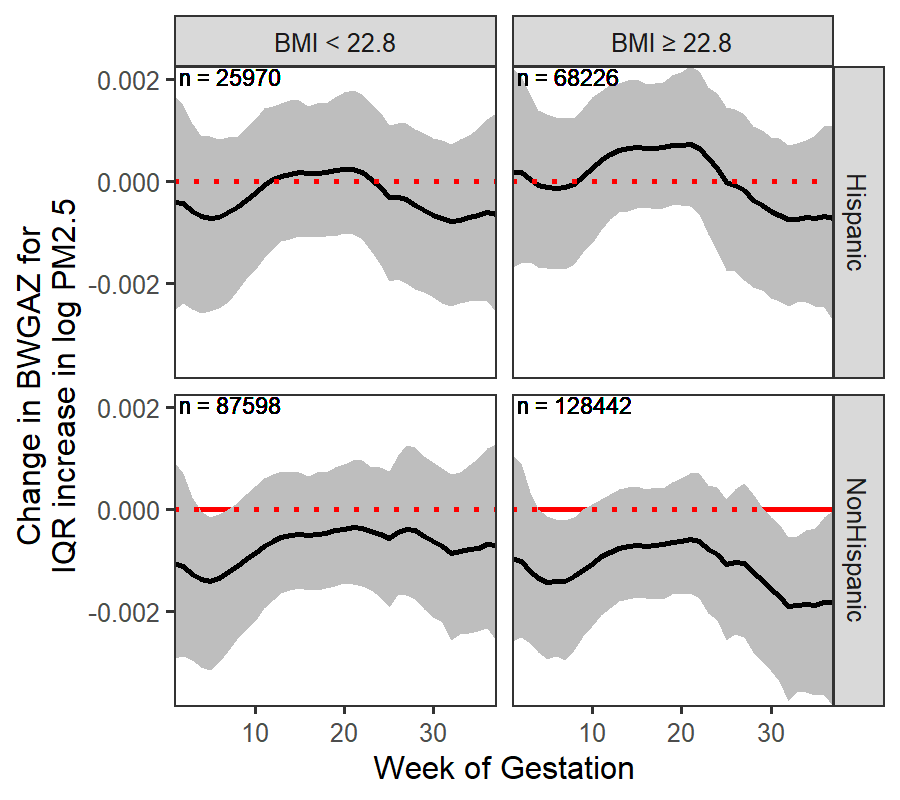}
        \vspace{0.25cm}
        \caption{}
        \label{fig:grp_dlm_hisp_bmi}
    \end{subfigure}
    \begin{subfigure}{0.49\textwidth}
        \includegraphics[height=6.25cm]{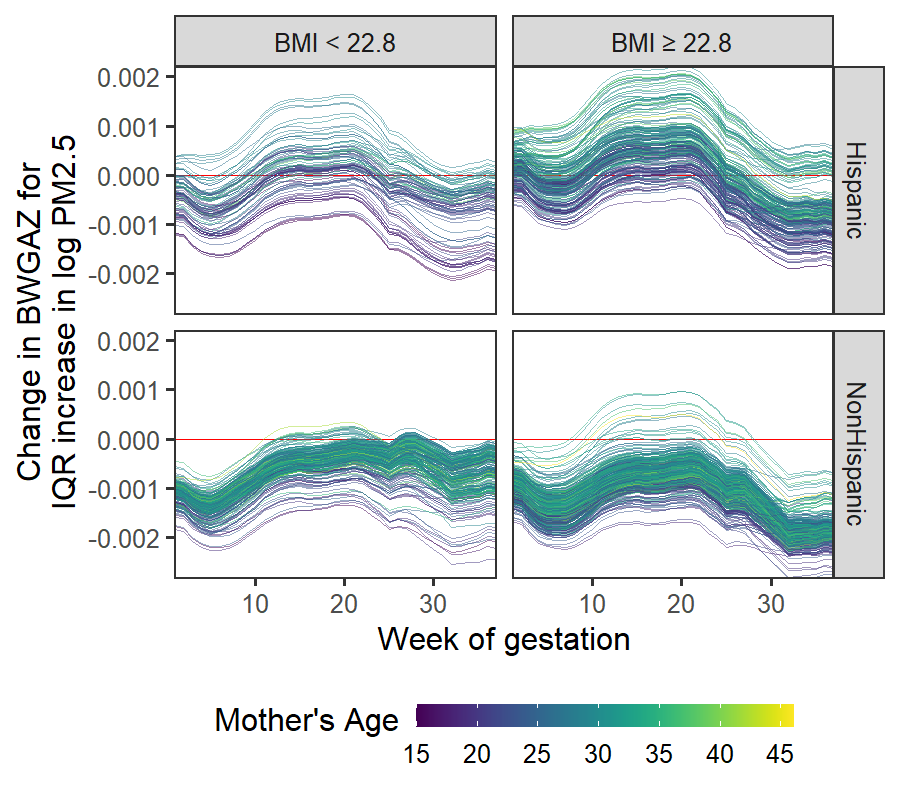}
        \caption{}
        \label{fig:ind_dlm_hisp_bmi}
    \end{subfigure}
    \caption{Panel (a) shows subgroup-specific DLM estimates with 95\% credible intervals. Panel (b) shows DLM estimates for 1,000 observations from our data analysis.  Both panels are grouped by Hispanic designation (rows) and BMI above/below 22.8 (columns). Panel (b) DLMs are colored according to a Mother's age at conception with lighter color representing older observations.}
\end{figure}

\begin{figure}[!ht]
\spacingset{1}
    \begin{subfigure}{.98\textwidth}
    \centering
        \includegraphics[height=6.5cm]{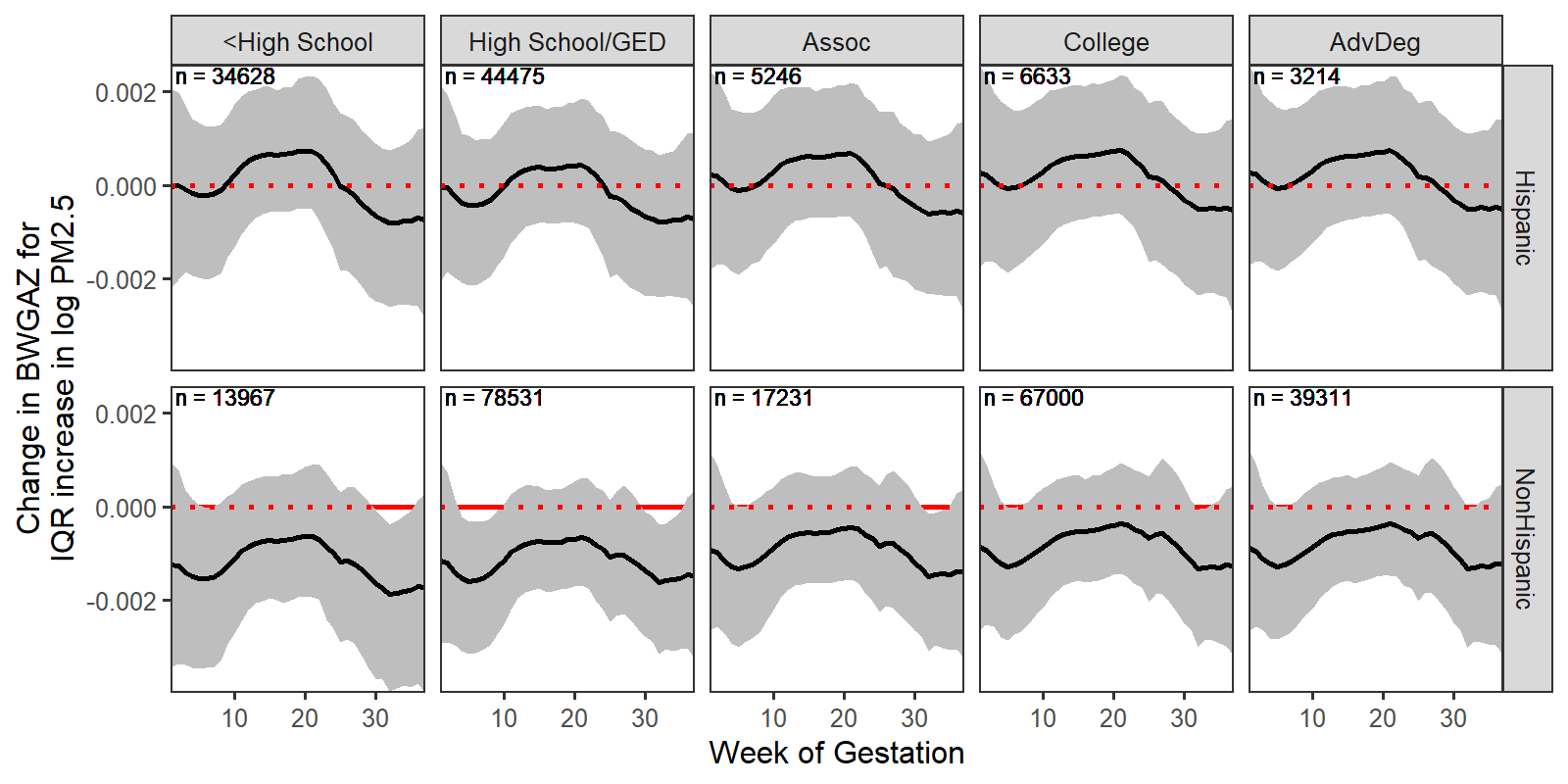}
        \caption{}
        \label{fig:grp_dlm_hisp_educ}
    \end{subfigure}
    \begin{subfigure}{.98\textwidth}
    \centering
        \includegraphics[height=6.5cm]{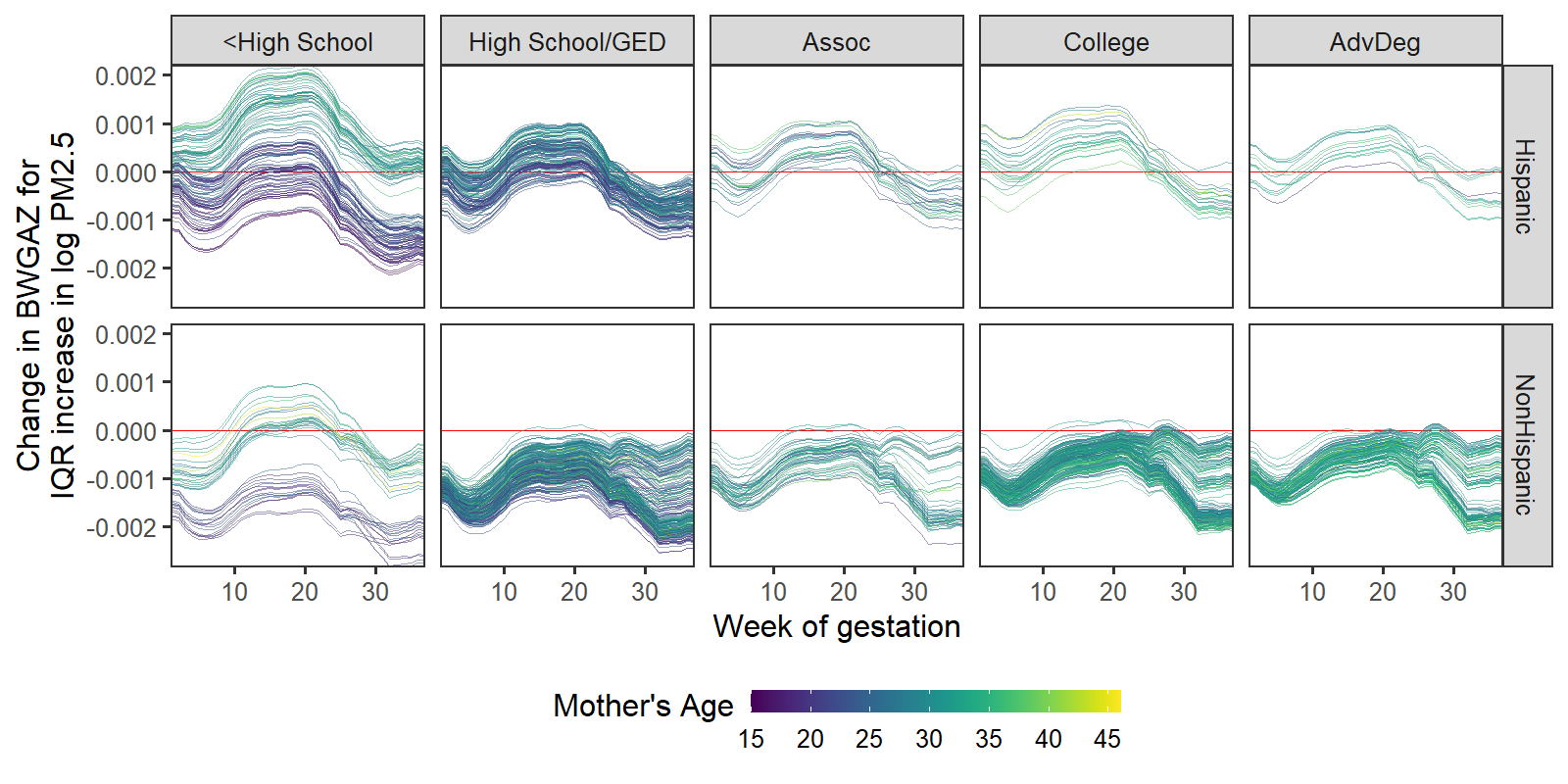}
        \caption{}
        \label{fig:ind_dlm_hisp_educ}
    \end{subfigure}
    \caption{Panel (a) shows subgroup-specific DLM estimates with 95\% credible intervals, divided by Hispanic and education modifiers. Panel (b) shows DLM estimates for 1,000 observations from our data analysis divided by Hispanic and education modifiers and colored by age, with lighter color representing older women.}
\end{figure}

Figures \ref{fig:grp_dlm_hisp_bmi} and \ref{fig:grp_dlm_hisp_educ} compare two-way subgroup-specific average distributed lag function estimates based on subgroup pairs with large modifier PIPs (Hispanic---BMI and Education---BMI). More complex multi-way interactions are presented in Supplementary Material Section 4. A consistent theme across subgroup analyses is a differential effect for Hispanic and non-Hispanic subgroups. We see a consistent negative exposure effect at all time points for the non-Hispanic groups and an early and late critical window that is present in many non-Hispanic subgroups. For Hispanic subgroups the exposure effect hovers around zero, which is generally consistent with no exposure effect.

Figure \ref{fig:grp_dlm_hisp_bmi} offers evidence of early-gestation susceptibility among all non-Hispanics. For non-Hispanics with BMI above 22.8, we detect a second critical window later in gestation during which increased PM$_{2.5}$ exposure is associated with lower birth weight. We find differences in critical windows among non-Hispanics based on level of education (Figure \ref{fig:grp_dlm_hisp_educ}). Non-Hispanics with less than a college degree show a late-gestation critical window and having at most a high school education suggests the presence of an early-gestation critical window. For Hispanics, there is no evidence of association at any education level.

%======================================
\subsection{Personalized distributed lag function estimates}
%======================================
The subgroup analysis in Section \ref{sec:subgroup-analysis} highlights broad trends across modifiers and identifies several vulnerable subgroups: non-Hispanic, BMI above 22.8, and less than college education. However, so far we have considered only two variables at a time. In reality, the observations in these subgroups vary across the entire observed range of the other modifiers considered in this analysis. In this section, we explore remaining variability in the distributed lag function among a sample of observations within these subgroups. Posterior analysis of individualized DLMs comes at high computational cost, hence we randomly select 1,000 observations and estimate their DLM based on the full data posterior samples.

Figures \ref{fig:ind_dlm_hisp_bmi} and \ref{fig:ind_dlm_hisp_educ} show individualized DLMs grouped according to the subgroup analysis in Section \ref{sec:subgroup-analysis} and colored to highlight differences in the distributed lag function according to a third modifier. This third modifier was selected based on observed variation in distributed lag function; however, other modifiers are also responsible for these differences.

Figure \ref{fig:ind_dlm_hisp_bmi} colors the estimated distributed lag function based on maternal age with darker color representing younger mothers. We see a trend towards younger observations having a larger negative effect. This difference by age is more noticeable for Hispanics with younger observations showing a more consistently negative effect. Figure \ref{fig:ind_dlm_hisp_educ} visualizes differences by education and Hispanic designation with color again representing continuous modification by maternal age. The distributed lag functions show pronounced differences between younger and older observations with less than a high school degree. Compared to their older counterparts, young Hispanic and non-Hispanic women with less than a high school education have consistently larger negative effects. Younger women with less education appear to be a highly vulnerable group to PM$_{2.5}$ exposure.

\subsection{Cumulative effect estimates and four-way interactions}

\begin{figure}[!ht]
\spacingset{1}
    \centering
    \includegraphics[height=7cm]{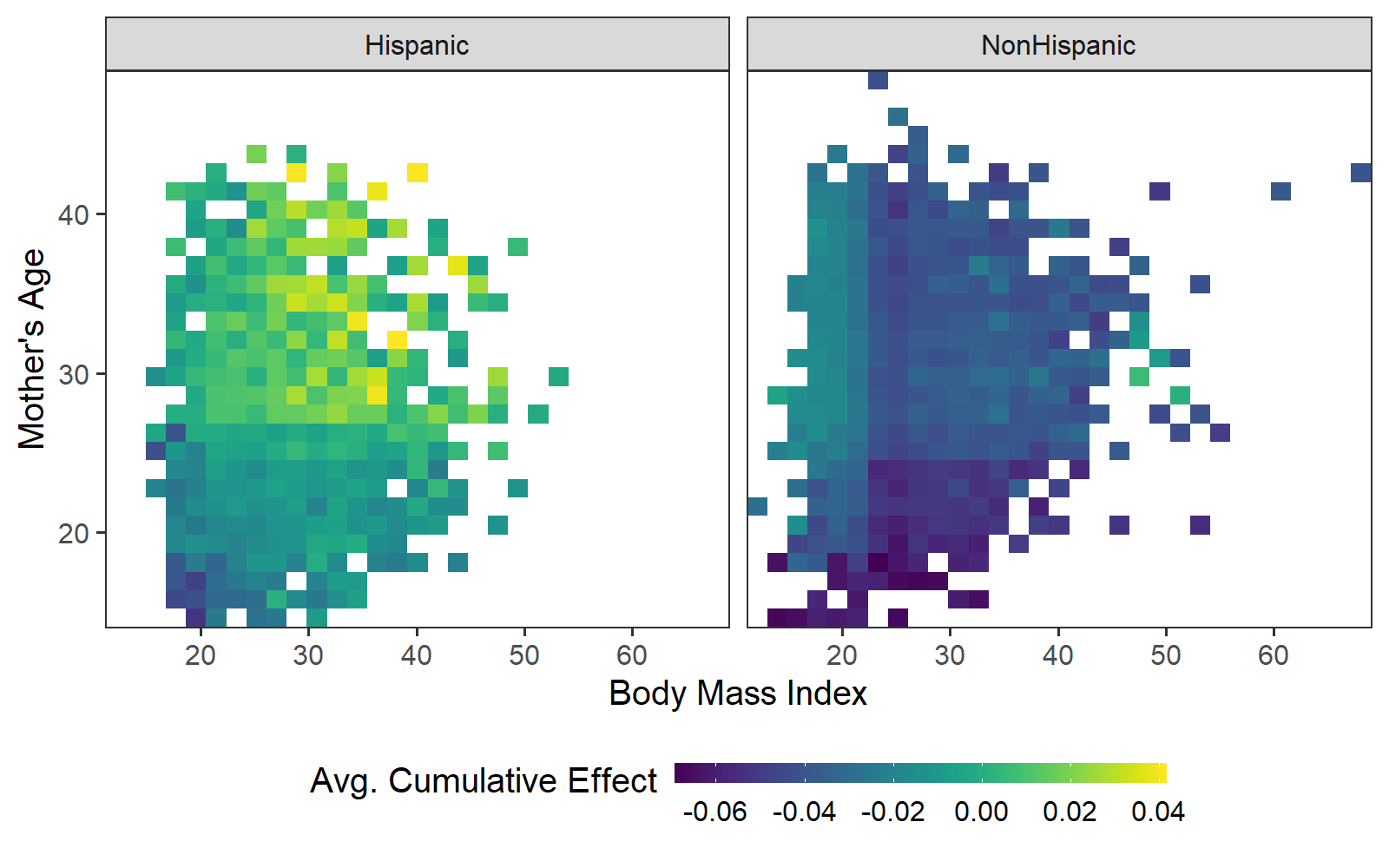}
    \caption{Heat map of average cumulative exposure effect for 5,000 observations from the data analysis. Each colored block is the average cumulative effect for observations with a particular body mass index (x-axis), age (y-axis) and Hispanic designation (panels). Darker color indicates a larger negative effect on BWGAZ associated with an IQR increase in PM$_{2.5}$ exposure throughout pregnancy.}
    \label{fig:avg_cumulative_effect}
\end{figure}

We next explored the differences in total exposure susceptibility across three modifiers simultaneously: maternal BMI, age, and Hispanic designation. We randomly selected 5,000 observations within our full data analysis and used the posterior samples to calculate their cumulative effect, or the sum of week-specific effects associated with an IQR increase in PM$_{2.5}$ exposure throughout pregnancy. The estimated cumulative effect was averaged across bins of the continuous modifiers, BMI and age. Results are visualized in Figure \ref{fig:avg_cumulative_effect}. For non-Hispanic observations, higher BMI and lower age correspond to the larger negative effects. Observations who are older and have lower BMI had the smallest cumulative effect of PM$_{2.5}$. Consistently across the non-Hispanic subgroup, increased exposure was related to lower BWGAZ. For Hispanic observations, the cumulative effect of exposure was centered around zero. We see evidence of a larger negative effect of exposure for younger Hispanic women; however this effect is less than the cumulative effect for non-Hispanics of the same age and BMI. There is little evidence of modification by BMI for Hispanics. 

The cumulative effect of PM$_{2.5}$ exposure on BWGAZ for non-Hispanic observations in this sample ranged from $-0.079$ to $0.007$, while for Hispanic observations the cumulative effect ranged from $-0.052$ to $0.051$. In the context of birth weight, the cumulative effect of an IQR increase in PM$_{2.5}$ exposure ranges from $-34.3$g to $3.2$g for non-Hispanic observations in the sample and from $-22.8$g to $22.1$g for Hispanic observations. These ranges are approximate because BWGAZ is adjusted for sex and gestational age.

%======================================
\section{Discussion}
%======================================
In this paper we analyzed an administrative birth cohort with the goal of understanding how individual-level modifiers alter the association between weekly average PM$_{2.5}$ exposure during the first 37 weeks of pregnancy and birth weight. By leveraging a large cohort with rich individual level exposure and covariate information, we were able to achieve a precision environmental health assessment that has not been previously possible due to small sample size, lack of individual-level data, and inadequate statistical methodologies. To enable our first-of-its-kind analysis, we proposed an innovative statistical learning framework combining the best available methodologies in distributed lag models and Bayesian additive regression trees. We demonstrated in a detailed simulation study that our HDLM methods estimate personalized critical windows and effect sizes with high precision and perform modifier variable selection.

In our analysis, we identified several subgroups with increased vulnerability to PM$_{2.5}$ exposure. These subgroups include non-Hispanic mothers who are younger or have higher BMI as well as non-Hispanic mothers with lower educational attainment. We estimated critical windows for these subgroups finding weeks during early- and late-gestation to be time periods where increased PM$_{2.5}$ exposure has a negative relationship with lower birth weights. There is ample research regarding the negative association between prenatal PM$_{2.5}$ exposure and birth weight \citep{Stieb2012, Lamichhane2015AOutcomes}, including identifying late term critical windows \citep{Sun2016TheMeta-analysis}, but there is limited comparable research into effect modification of air pollution on birth weight. \citet{Lakshmanan2015AssociationsIndex} found PM$_{2.5}$ to be associated with lower birth weight for males with obese mothers (BMI $\geq30$) and \citet{Bell2007AmbientMassachusetts} found larger effects of prenatal PM$_{2.5}$ exposure on birth weight for Black mothers, however these studies used average exposure during pregnancy. No study to our knowledge has considered racial or ethnic disparities in regard to air pollution exposure critical windows for birth weight. One possible explanation for the limited effect of PM$_{2.5}$ on birth weight for Hispanics in our study is the immigrant birth weight paradox \citep{Chu2022ThePollution}, where foreign-born women have higher birth weight children despite lower socioeconomic indicators and less prenatal care. All of the aforementioned studies used only a few predefined subgroups that were hypothesized by the researchers to show effect heterogeneity. In contrast, our analysis allows for a data driven approach to identifying the individual characteristics responsible for driving effect modification.

HDLM enabled us to go beyond a standard stratified analysis and observe exposure-outcome relationships at the individual level. In particular, we found that select observations had much larger critical windows than suggested by the population average effect, while others showed little evidence of exposure effects. Likewise, cumulative effect estimates showed a range of variability in the total change in birth weight related to increased PM$_{2.5}$ exposure. For example, compared to the population average, some observations (non-Hispanic, young, high BMI) had a 3 times larger decrease in birth weight associated with an IQR increase in PM$_{2.5}$ exposure across pregnancy. The personalized exposure-response function provides stakeholders with detailed information regarding the burden of air pollution on health and reiterates the need for air pollution standards that protect the most vulnerable populations.

The present case study allowed for a high-powered analysis by combining the gold standard of a large sample size and ample individual level variables. With regard to data quality, we anticipate a high degree of accuracy tied to the reporting of birth certificate data due to the involvement of medical professionals. The ambient PM$_{2.5}$ exposure measurements at the census tract level have high predictive accuracy \citep{Berrocal2010AModels}, but spatial misalignment between exposures and observations across a census tract may lead to slight bias in the effect estimates \citep{Gryparis2009MeasurementEpidemiology}. By including spatial, temporal, and meteorological covariates (county, elevation, year, month, temperature) we are able to partially account for unmeasured confounding correlated with location and time of birth \citep{Lu2007OnEpidemiology}. Although we are not able to account for residential mobility of some mothers during pregnancy, research shows this does not bias the results \citep{Warren2018InvestigatingPregnancy}. Under several additional assumptions such as no unmeasured confounding, consistency of treatment, positivity, no interference, and ignorable exposure assignments under a potential outcomes framework, one might argue existence of a causal relationship for the conditional average treatment effects based on any particular level of modifiers. As is true in any causal inference setting, the majority of these assumptions cannot be tested. In any case, the present analysis and results could be fortified by including a more flexible confounding model.

Our case study and simulations showcase the bias incurred from a standard DLM analysis that does not allow for effect heterogeneity when it exists. To address these shortfalls we propose HDLM as a data-driven method to discover modifiers responsible for heterogeneity in an exposure-time-response relationship. The results highlight the need for methodologies that account for individual differences. Ultimately our method and findings can lead to personalized environmental health decision making and pinpoint at-risk individuals for targeted public health interventions. 

\section*{Supplementary Materials}
\textbf{Supplement:} Technical derivations, additional simulation and data analysis results.\\
\textbf{R-package dlmtree:} Code to implement the software in this manuscript is publicly available at \url{https://github.com/danielmork/dlmtree}. Code to reproduce simulation and data analysis figures and results is provided at \url{https://github.com/danielmork/HDLM}.

\section*{Acknowledgements}
This work was supported by National Institutes of Health grants ES029943, ES028811, and P30-ES000002. This research was also supported by USEPA grants RD-839278 and RD-83587201. Its contents are solely the responsibility of the grantee and do not necessarily represent the official views of the USEPA. Further, USEPA does not endorse the purchase of any commercial products or services mentioned in the publication. This work utilized the RMACC Summit supercomputer, which is supported by the National Science Foundation (awards ACI-1532235 and ACI-1532236), the University of Colorado Boulder and Colorado State University. The RMACC Summit supercomputer is a joint effort of the University of Colorado Boulder and Colorado State University. These data were supplied by the Center for Health and Environmental Data Vital Statistics Program of the Colorado Department of Public Health and Environment, which specifically disclaims responsibility for any analyses, interpretations, or conclusions it has not provided.

\section*{Conflict of interest}
No potential competing interest was reported by the authors.

%======================================
\bibliography{references}
%======================================
\end{document}